\documentclass[10pt,draft]{dis03}
\usepackage{epsf,amsmath}

\begin{document}

\title{Charm dijet angular distributions \\
in $\gamma p$ collisions with ZEUS at HERA
}

\author{Leonid~Gladilin
\thanks{On leave from Moscow State University,
supported by the U.S.-Israel BSF} \\
(on behalf of the ZEUS Collaboration) \\
DESY, ZEUS experiment, Notkestr. 85, 22607 Hamburg, Germany \\
E-mail: gladilin@mail.desy.de}

\maketitle

\begin{abstract}
\noindent
Dijet angular distributions of photoproduction events in which
a $D^{*\pm}$ meson is produced in association with
one of two energetic jets have been measured using an integrated
luminosity of $120\,$pb$^{-1}$.
The results are compared with predictions from leading-order
parton-shower Monte Carlo models and with
next-to-leading-order QCD calculations.
\end{abstract}

\section{Introduction}
In photoproduction processes at HERA, a quasi-real photon
with virtuality $Q^2 \sim 0$ is emitted by the incoming
electron and interacts with the proton.
At leading order (LO) in QCD,
two types of processes are responsible for the production 
of heavy quarks:
the direct photon processes,
where the photon participates as a point-like particle,
and the resolved photon processes, where
the photon acts as a source of partons.
In the direct photon-gluon-fusion (PGF) process
(Fig.~1a), the entire photon momentum is involved
and the propagator is a spin-$\frac{1}{2}$ quark propagator.
In resolved photon processes, a parton from the photon
scatters off a parton from the proton (Figs.~1b, 1c and 1d),
and only a fraction of the photon momentum participates
in the hard scatter.
Charm quarks present in the parton distributions of the photon, as
well as of the proton, lead to processes
like $cg~\to~cg$,
which are called charm-excitation processes (Figs.~1c and 1d).
The dominant $t$-channel charm-excitation diagram has
a spin-$1$ gluon propagator (Fig.~1d).
At next-to-leading order (NLO) in QCD, only the sum of direct and resolved
processes is unambiguously defined. 

To identify contributions of different charm-production
processes and to verify relevant QCD calculations,
photoproduction of events with a $D^{*\pm}$ meson and at least
two energetic jets has been studied~\cite{dstar,cjets}.

\begin{figure}[!thb]
\vspace*{11.cm}
\begin{center}
\includegraphics{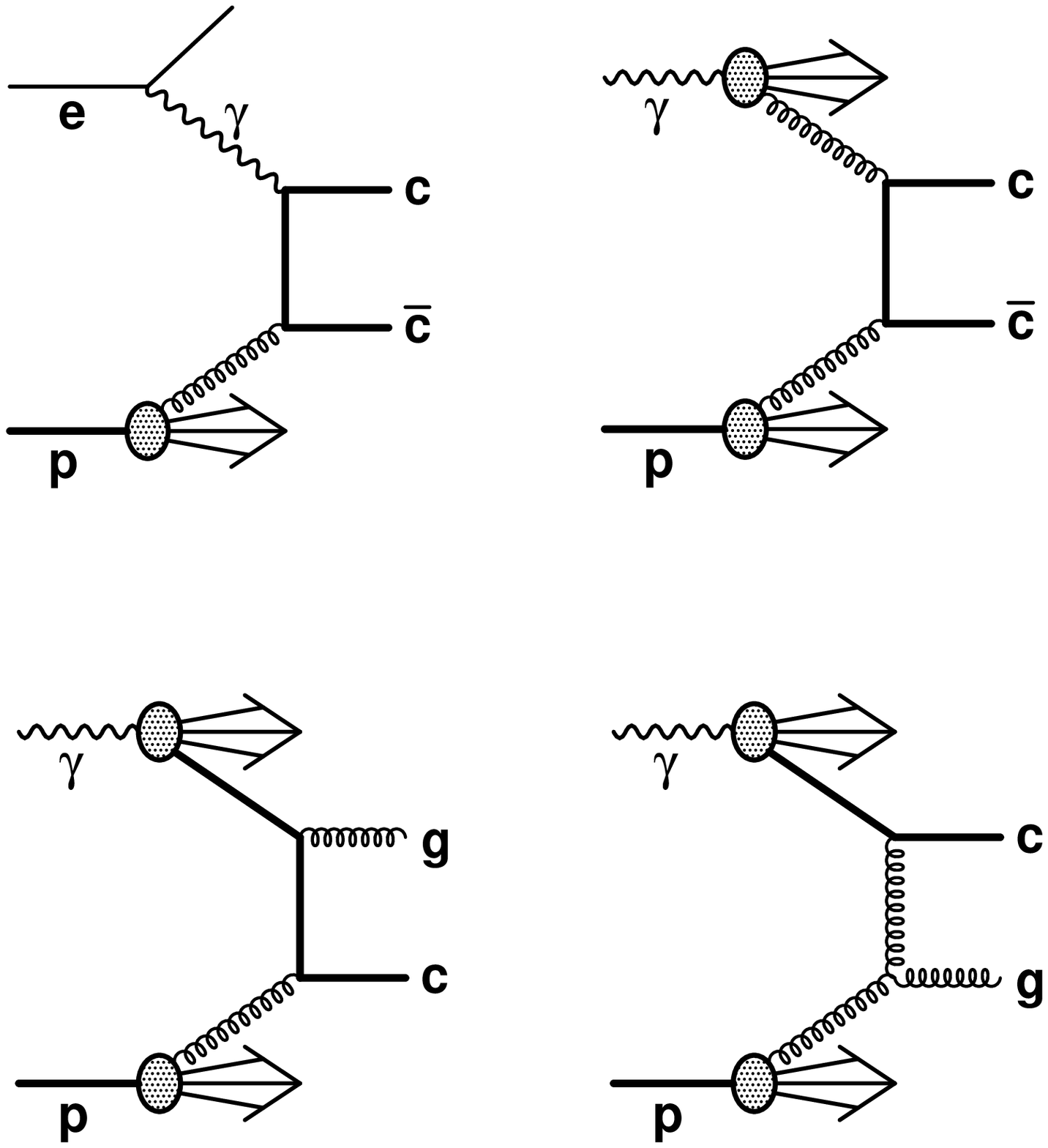}
\vspace*{-2.0cm}
  \vskip-5.1cm
  \centerline{\large\bf\kern3.0cm (a)\kern3.8cm (b)\hfill}
  \vskip4.1cm
  \centerline{\large\bf\kern3.0cm (c)\kern3.8cm (d)\hfill}
\caption[*]{
LO QCD charm-production diagrams:
(a) direct photon: $\gamma g~\to~c{\bar c}$;
(b) resolved photon: $g g~\to~c{\bar c}$;
(c) resolved-photon charm excitation:
$cg~\to~cg$ ($u$-channel);
(d) resolved-photon charm excitation:
$cg~\to~cg$ ($t$-channel).
}
\end{center}
\end{figure}

\section{QCD calculations of charm photoproduction}
The Monte Carlo (MC) programs PYTHIA~\cite{pythia}
and HERWIG~\cite{herwig} simulate heavy-quark photoproduction
in the framework of the collinear approach using
the on-shell LO matrix elements for direct and resolved photon processes
(including charm excitation).
Higher-order QCD effects are simulated in the leading-logarithmic
approximation with initial- and final-state radiation obeying
DGLAP evolution.

The MC program CASCADE~\cite{cascade} simulates heavy-quark
photoproduction in the framework of the semi-hard or $k_t$-factorisation
approach using the off-shell LO PGF matrix element.
The resolved photon processes are reproduced by the initial-state
radiation based on CCFM evolution.

The NLO QCD calculations of differential cross sections for
photoproduction of charm dijet events are available~\cite{fmnr} in
the fixed-order scheme assuming no explicit
charm-excitation component.
Charm photoproduction cross sections are calculated
in the framework of the collinear approach using
the on-shell matrix elements.

\section{Dijet angular distributions in photoproduction of charm}
An experimental separation of the direct and resolved processes was
obtained
by a selection on the variable
$$x_\gamma^{\rm obs} =
\frac{E_T^{\rm jet1}\eta^{\rm jet1}+E_T^{\rm jet2}\eta^{\rm jet2}}
{2yE_e},$$
where $yE_e$ is the initial photon energy.
The variable $x_\gamma^{\rm obs}$ is the fraction of the photon's
momentum contributing to the production of the two jets.
The measured $D^{*\pm}$ photoproduction cross section peaks
at $x_\gamma^{\rm obs}\sim 1$, in agreement with the expectation
for direct photon processes. A large cross section is also measured 
at low $x_\gamma^{\rm obs}$, where resolved processes are expected
to contribute significantly.
The selection of $x_\gamma^{\rm obs}>0.75$ and $x_\gamma^{\rm obs}<0.75$
was used to obtain samples enriched in direct and resolved
photon processes, respectively.

\begin{figure}[!thb]
\vspace*{8.5cm}
\begin{center}
\includegraphics{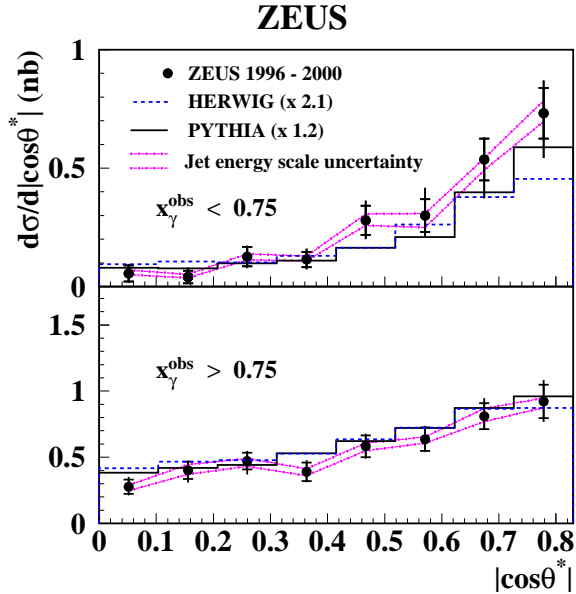}
\caption[*]{
Differential cross-section $d\sigma/d |\cos\theta^*|$ for the data
compared with PYTHIA and HERWIG MC simulations.
Results are given separately for samples enriched in
resolved photon events (upper) and for samples enriched in
direct photon events (lower).
}
\end{center}
\end{figure}

Sensitivity to the spin of the propagator in the hard subprocess
was obtained by measuring
the angle between the jet-jet axis and the beam axis in the dijet rest frame.
This angle, $\theta^*$, was reconstructed using
\begin{equation}
\cos \theta^* = \tanh\left(\frac{\eta^{\rm jet1}-\eta^{\rm jet2}} {2}\right).
\end{equation}
The angular dependence of the cross section for processes
with a spin-1 gluon propagator is approximately
$\propto(1-|\cos\theta^*|)^{-2}$. This cross section rises more
steeply with increasing $|\cos\theta^*|$ than that for processes
with a spin-$\frac{1}{2}$ quark propagator, where the angular
dependence is approximately $\propto(1-|\cos\theta^*|)^{-1}$.

Figure~2 shows the differential cross sections as a function
of $|\cos\theta^*|$ for the resolved- and  direct-enriched samples.
The angular distribution of resolved-enriched events
exhibits a more rapid rise towards
high values of $|\cos\theta^*|$ than does the distribution
of direct-enriched events.
This observation suggests a large contribution from
the $t$-channel charm-excitation diagram with
a spin-$1$ gluon propagator (Fig.~1d).
The data are compared to the PYTHIA and HERWIG predictions
normalized to the data. The MC predictions
provide adequate descriptions of the shapes of the data distributions.

\begin{figure}[!thb]
\vspace*{8.5cm}
\begin{center}
\includegraphics{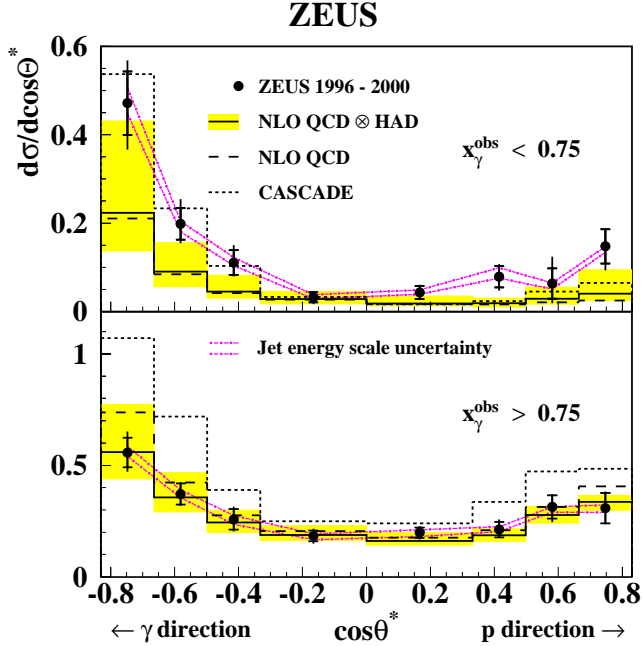}
\caption[*]{
Differential cross-section $d\sigma/d \cos\theta^*$ for the data
compared with NLO QCD predictions.
Results are given separately for samples enriched in
resolved photon events (upper) and for samples enriched in
direct photon events (lower).
}
\end{center}
\end{figure}

The two jets can be distinguished by associating the $D^{*\pm}$ meson
to the closest jet in $\eta-\phi$ space and calling this
jet 1 in Eq.~(1).
Figure~3 shows the differential cross sections as a function
of $\cos\theta^*$ for the resolved- and  direct-enriched samples.
The angular distribution of resolved-enriched events
exhibits a large asymmetry
with a mild rise towards $\cos\theta^*=1$ (proton direction) and
a strong rise towards
$\cos\theta^*=-1$ (photon direction).
This observation shows that dijet events
with $x_\gamma^{\rm obs}<0.75$ are dominantly produced by charm quarks
coming from the photon side.
The $\cos\theta^*$ distribution of direct-enriched events 
is almost symmetric, as expected for the PGF process.
A slight asymmetry can be explained by
the feedthrough from resolved photon processes near
$\cos\theta^*=-1$~\cite{cjets}.

The CASCADE and NLO predictions are compared to the data in Fig.~3.
For $x_\gamma^{\rm obs}<0.75$,
the CASCADE prediction describes the data
in the photon direction and underestimates the data in the proton direction.
For $x_\gamma^{\rm obs}>0.75$,
the prediction overestimates the data in all regions of $\cos\theta^*$,
although the shape is described reasonably well.
The NLO prediction is below the data for $x_\gamma^{\rm obs}<0.75$
and in agreement with the data for $x_\gamma^{\rm obs}>0.75$.
The shapes are reasonably well described by the NLO predictions.

\section{Summary}
The measured charm dijet angular distributions
show that dijet events at low $x_\gamma^{\rm obs}$
are dominantly produced by charm quarks coming from the photon side.
The shapes of the measured distributions are described adequately
by the LO collinear calculations (PYTHIA, HERWIG) including
a large charm-excitation component.

The shapes of the distributions are reasonably well described by
the NLO collinear calculations in the fixed-order
scheme assuming no explicit charm-excitation component.
The LO $k_t$-factorisation
calculations with the initial-state radiation
based on the CCFM evolution (CASCADE) also reproduce
the shapes reasonably well.
However, the NLO and CASCADE calculations do not reproduce relative
contributions of charm dijet events with high and low $x_\gamma^{\rm obs}$
values. The NLO prediction is below the data at low $x_\gamma^{\rm obs}$
and the CASCADE prediction overestimates the data at high $x_\gamma^{\rm obs}$.

\end{document}